\documentclass[aps,prc,twocolumn,floatfix,showpacs,nofootinbib,amsmath,amssymb]{revtex4-1}
\usepackage{graphicx}
\usepackage{dcolumn}
\usepackage{bm}
\usepackage{color}
\newcommand{\be}{\begin{equation}}
\newcommand{\ee}{\end{equation}}
\newcommand{\ba}{\begin{eqnarray}}
\newcommand{\ea}{\end{eqnarray}}
\newcommand{\bd}{\begin{displaymath}}
\newcommand{\ed}{\end{displaymath}}
\newcommand{\bea}{\begin{eqnarray}}
\newcommand{\eea}{\end{eqnarray}}

\begin{document}
\title{Neutron stars with a crossover equation of state}

\author{J. I. Kapusta and T. Welle}
\smallskip

\affiliation{School of Physics and Astronomy, University of Minnesota, Minneapolis, Minnesota 55455 USA}

\begin{abstract}
The question of whether quark matter exists in neutron stars is a long standing one.  Generally one finds that a first order phase transition from baryons to quarks softens the equation of state so much that the star would collapse into a black hole.  We consider a crossover equation of state, similar to the crossover that is found in lattice QCD studies at finite temperature and zero or small baryon chemical potentials.  We find that with reasonable parameters it may be possible to support neutron stars up to about 2.2 solar masses.  In that case 1 to 10\% of the pressure would be contributed by quark matter in the central core of the highest mass stars.
\end{abstract}
\date{\today}

\maketitle

%%%%%%%%%%%%%%%%%%%%%%%%%%%%%%%%%%%

Neutron stars are an interesting astrophysical laboratory to study strong gravitational fields and the properties of matter at the highest attainable baryon densities in the universe.  The LIGO/Virgo collaboration has observed gravitational waves from the merger of neutron stars \cite{LV}.  Such mergers are predicted to have produced many of the heavy elements in the universe \cite{heavies}.  The central baryon densities are theoretically estimated to be about ten times greater than in atomic nuclei.  Such baryon densities will likely be reached or exceeded in the Beam Energy Scan II at RHIC (Relativistic Heavy Ion Collider) \cite{QMseries}.  The advantage of heavy ion collision experiments is that they are measurable and repeatable on earth.  Their disadvantage is that the volume in which these densities are achieved is transient and involve a length scale of 10 fm, whereas the length scale in the centers of neutron stars is 1 km, a difference of 17 orders of magnitude.

For a long time it was thought that neutron stars might have an upper limit of about 1.5 to 1.6 solar masses $M_{\odot}$ \cite{LP1}.  There were a few with larger masses but they had significant uncertainities.  Then stars with masses 1.97$\pm 0.04$ $M_{\odot}$ \cite{Demorest} (subsequently updated to 1.928$\pm 0.017$ $M_{\odot}$ \cite{Fonseca}), 2.01$\pm 0.04$ $M_{\odot}$ \cite{Antoniadis}, and 2.14$\pm 0.10$ $M_{\odot}$ \cite{Cromartie} were discovered.  These require substantial interaction pressure for them to be held up from gravitational collapse.  For comparison, an equation of state with relativistic but noninteracting protons, neutrons, and electrons can only support a 0.7 solar mass star.

The equation of state for dense matter relevant to neutron stars has been studied for more than fifty years.  There exists a vast literature which we do not intend to review here.  One approach is relativistic mean field theory \cite{Serot-Walecka,Kapusta-Gale} .  In this approach baryons move independently in the mean scalar and vector meson fields generated by the baryons themselves.  Although not entirely satisfactory for several reasons, it has the advantage that the known properties of nuclear and neutron matter near normal nuclear density $n_0 = 0.153$ fm$^{-3}$ can be fit with the parameters in the Lagrangian, and then the equation of state can be extrapolated to higher densities while respecting special relativity and thermodynamic relationships.  As the density and therefore chemical potential increase it becomes thermodynamically favorable to form hyperons.  But the strength of the interactions between hyperons and nucleons and among hyperons is poorly constrained by hypernuclei, which is true of all theoretical approaches.  Adding degrees of freedom softens the equation of state, meaning that at the same energy density the pressure is decreased.  This makes it difficult to make massive neutron stars with most choices of coupling constants \cite{JEllis}.  Reaching at least 2 $M_{\odot}$ requires parameter choices that effectively eliminate hyperons \cite{SU3}.  A first order phase transition to quark matter would also soften the equation of state since the quark phase has higher energy density than the hadronic phase at the same pressure.  Hence it is highly unlikely that a first order phase transition to quark matter could occur before the star would collapse to a black hole \cite{JEllis}.  A second order transition also softens the equation of state, but not by much \cite{JEllis}.  

Precision lattice QCD simulations at finite temperature and zero or small chemical potential clearly show a crossover transition from hadronic matter to quark--gluon plasma \cite{latticeQCD}.  The equation of state was accurately modeled in Ref. \cite{matchingpaper} using a switching function to transition smoothly from a hadronic gas to quark--gluon plasma.  Motivated by this and by the discussion of neutron stars above we entertain the possibility that the transition from cold nuclear or neutron matter to quark matter is also a crossover and study the general implications for neutron stars.  At $T=0$ the equation of state is given by
\be
P(\mu) = S(\mu) P_q(\mu) + (1 - S(\mu))  P_h(\mu)
\ee
Here $P_h(\mu)$ is the hadronic pressure and $P_q(\mu)$ is the quark pressure as functions of the baryon chemical potential $\mu$.  The $S(\mu)$ is a switching function which ranges between 0 and 1 as $\mu$ increases.  It must be infinitely differentiable so as not to introduce an artificial phase transition of any order.  Reference \cite{matchingpaper} suggests the form
\be
S(\mu) = \exp\left[ - (\mu_0/\mu)^4 \right]
\ee
with the reasonable estimate that $\mu_0 \approx 1.65$ GeV.  This parameter could not be precisely determined by comparison with the lattice results but simply as a baseline estimate.  This approach allows freedom in the choice of hadron and quark pressures.

Based on the previous discussion we shall, for simplicity and sake of illustration, use a relativistic mean field equation of state including neutrons interacting via scalar $\sigma$ and vector $\omega$ plus $\rho$ mesons.  The potential $U(\sigma)$ includes quadratic, cubic, and quartic terms.  This amounts to five independent parameters.  They are fit to the properties of normal isospin symmetric nuclear matter to reproduce the saturation density $n_0$, binding energy 16.3 MeV, Landau mass 0.83 $m_N$, compressibility $K = 250$ MeV, and symmetry energy $S = 32.5$ MeV.  See Ref. \cite{Kapusta-Gale} or other references for details.  The perturbative QCD quark matter pressure comes from Ref. \cite{quarkmatter}.  These include contributions up to and including $\alpha_s^2$.  The strong coupling $\alpha_s(\mu)$ satisfies the renormalization group equation using the 3--loop $\beta$ function.  The renormalization scale is taken from fits to the lattice QCD results as in Ref. \cite{matchingpaper}.  All quark masses are set to zero.  Thus both phases are electrically neutral without the need for electrons or muons.  The only free parameter is $\mu_0$ whose scale is already known.

\begin{figure}[h]
\includegraphics[width=1.05\columnwidth]{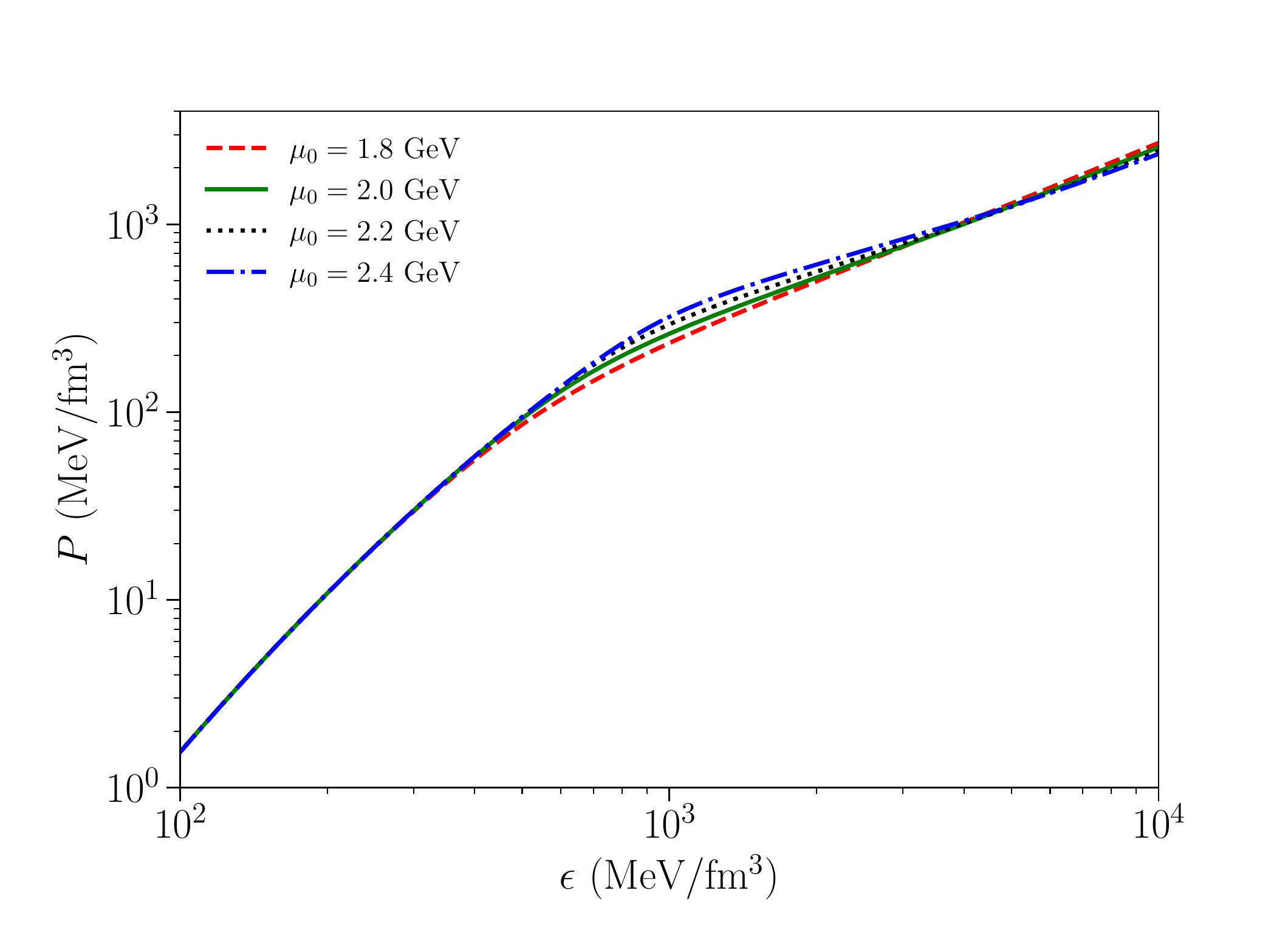}
\caption{Pressure versus energy density for different choices of $\mu_0$.}
\label{PvsE}
\end{figure}
Figure \ref{PvsE} shows the pressure versus energy density for four values of $\mu_0$.  The influence of quark matter becomes apparent between 0.5 and 1.0 GeV/fm$^3$ depending on the value of $\mu_0$.  On a log--log plot they all appear very similar.

\begin{figure}[h]
\includegraphics[width=1.05\columnwidth]{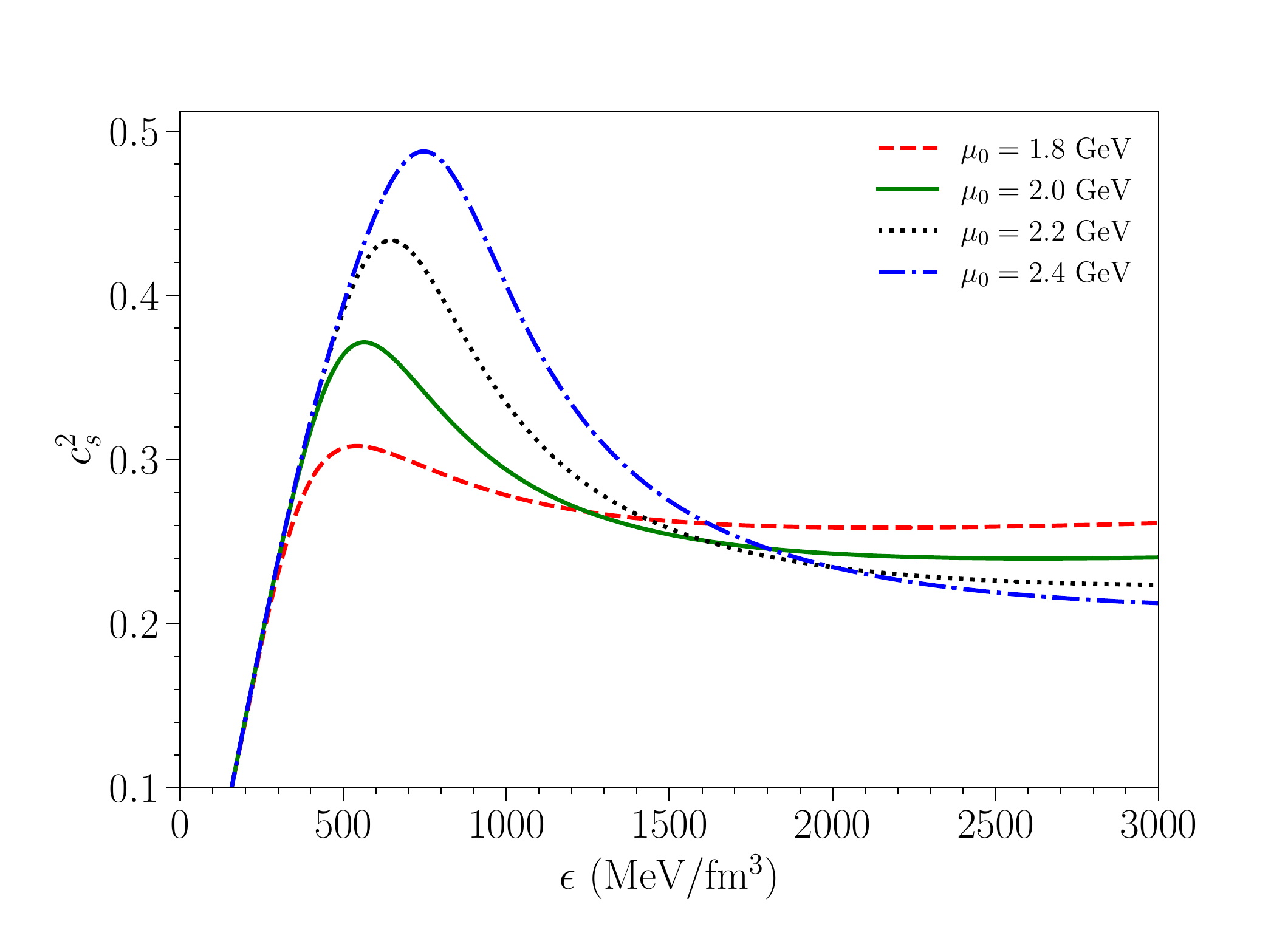}
\caption{Speed of sound squared versus energy density for different choices of $\mu_0$.}
\label{cs2}
\end{figure}

A more differential probe is the slope $dP/d\epsilon$.  At $T=0$ this is the adiabatic speed of sound squared $c_s^2$ and is plotted in Fig. \ref{cs2}.  This behavior can be understood as follows.  For the relativistic nuclear mean field theory the asymptotic behavior at large density or chemical potential is
\be
P \rightarrow \epsilon \rightarrow \frac{1}{2}\left[ \frac{g_{\omega}^2}{m_{\omega}^2} + \frac{g_{\rho}^2}{4 m_{\rho}^2}\right] n^2
\ee
where $g_{\omega}$ and $g_{\rho}$ are the baryon--vector meson coupling constants, $m_{\omega}$ and $m_{\rho}$ are the vector meson masses, and $n = dP/d\mu$ is the baryon density.  Hence $c_s^2 \rightarrow 1$ in the absence of quark matter.  Since QCD is asymptotically free $c_s^2 \rightarrow 1/3$ for quark matter, apart from logarithmic corrections.  With a crossover equation of state it is therefore natural that $c_s^2$ would at first increase with energy density, reach a maximum less than 1, and then decrease towards an asymptotic limit of 1/3.  To determine the behavior at asymptotically high density consider the quark pressure at two--loop order
\be
P_q = \frac{N_f}{4\pi^2} \left( \frac{\mu}{3} \right)^4 \left( 1 - \frac{2 \alpha_s}{\pi} \right)
\ee
where $N_f$ is the number of massless quarks.  Using the one--loop $\beta$ function this leads to
\be
c_s^2 = \frac{n}{\mu \, dn/d\mu} = \frac{1}{3} \left[ 1 - \frac{(33 - 2 N_f)}{9\pi^2}  \alpha_s^2 \right]
\ee
Hence the limit of 1/3 is reached from below.  The behavior shown in Fig. \ref{cs2} is qualitatively the same as speculated in Ref. \cite{Tews}.  It is also similar to the behavior of quarkyonic matter \cite{quarkyonic1} although the theoretical approaches appear to be very different, with the latter resulting in a second order phase transition.

It is interesting to examine the mathematical limit $\mu \gg \mu_0$.  In this limit $P_h \rightarrow \frac{1}{2}\left( \frac{g_{\omega}^2}{m_{\omega}^2} + \frac{g_{\rho}^2}{4 m_{\rho}^2}\right)^{-1} \mu^2$ and so
\ba
(1-S) P_h &\rightarrow& \frac{1}{2} \left( \frac{g_{\omega}^2}{m_{\omega}^2} + \frac{g_{\rho}^2}{4 m_{\rho}^2}\right)^{-1} \frac{\mu_0^4}{\mu^2} \rightarrow 0 \nonumber \\
S P_q &\rightarrow& \frac{N_f}{4\pi^2} \left( \frac{\mu}{3} \right)^4 \left( 1 - \frac{2 \alpha_s}{\pi} \right) - \frac{N_f \mu_0^4}{(18 \pi)^2}
\ea
plus a correction of order $\alpha_s$ in the last term above.  Thus there is a constant shift in the pressure at very high density; $\mu_0^4$ essentially acts like a bag constant.

\begin{figure}[h]
\includegraphics[width=1.05\columnwidth]{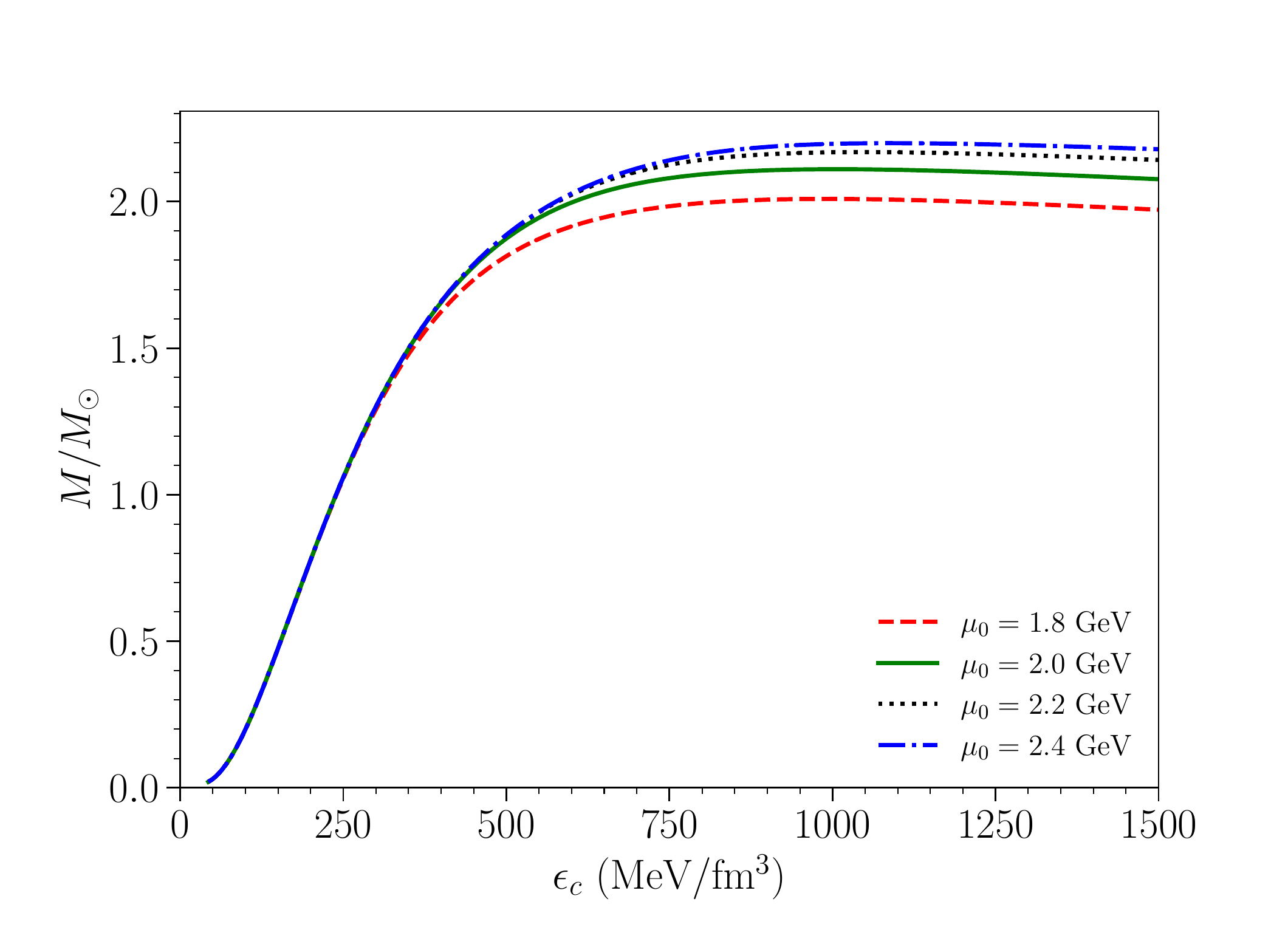}
\caption{Neutron star mass in units of solar mass versus central energy density for different choices of $\mu_0$.  The maximum masses are 2.01, 2.11, 2.17 and 2.20 $M_{\odot}$.}
\label{Mvseps}
\end{figure}

Figure \ref{Mvseps} shows the mass as a function of central energy density for the four choices of $\mu_0$.  The maximum mass increases with $\mu_0$ and with central energy density $\epsilon_c$, which is to be expected.  For any of these choices of $\mu_0$, the value of $\epsilon_c$ for the maximum mass star is beyond the peak in $c_s^2$, as can be seen by comparison of Figs. \ref{cs2} and \ref{Mvseps}.  The observation of 2.01$\pm$0.04 and 2.1$\pm$0.10 solar mass stars means that $\mu_0$ cannot be less than about 1.8 GeV, at least with our choice for $P_h$.

\begin{figure}[h]
\includegraphics[width=1.05\columnwidth]{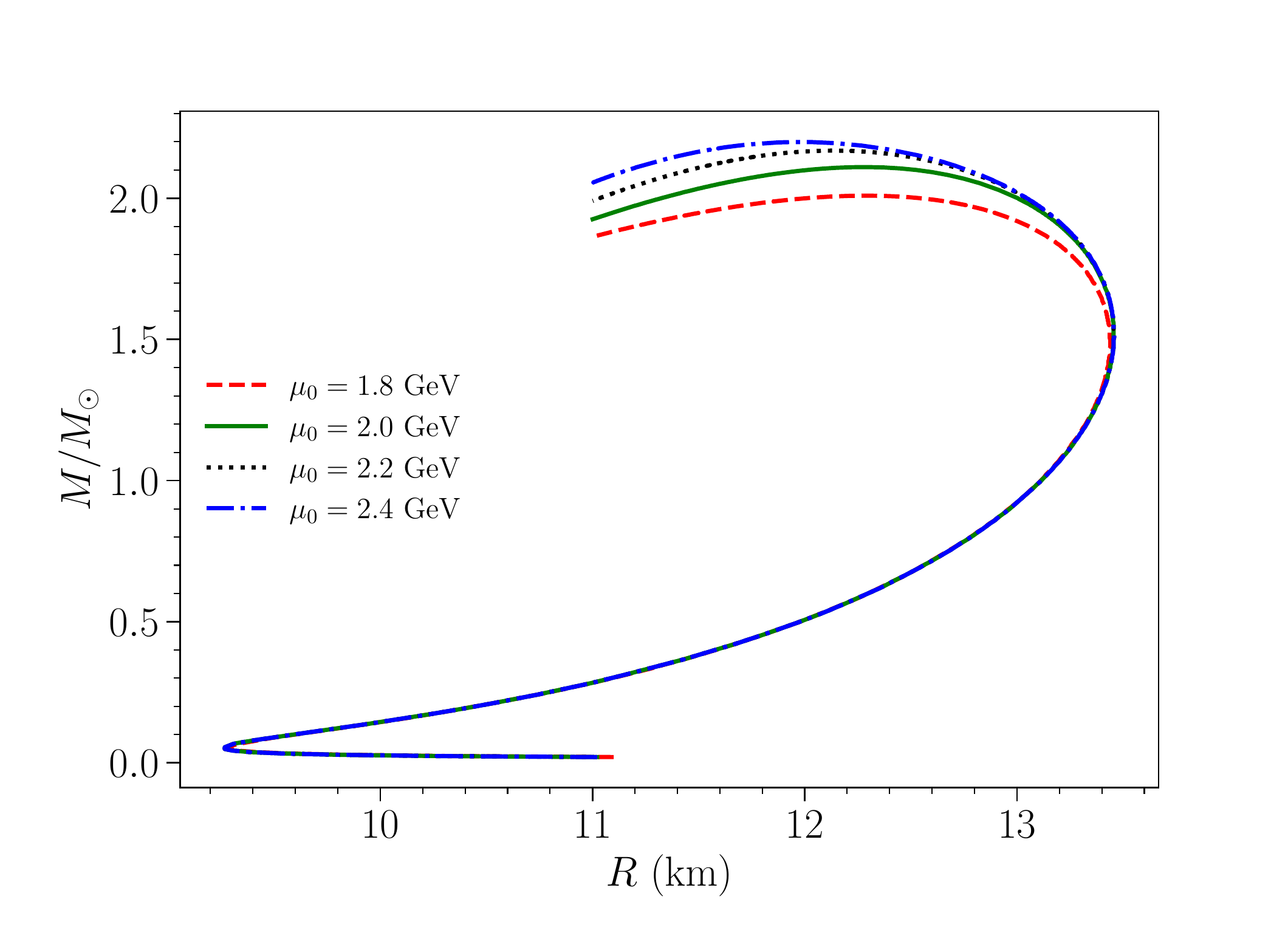}
\caption{Neutron star mass in units of solar mass versus surface radius for different choices of $\mu_0$. }
\label{MvsR}
\end{figure}

Figure \ref{MvsR} displays the mass--radius relationship. The overall shape of the curves is typical.  As $\mu_0$ increases so does the maximum mass since larger values favor the hadronic phase. Also, as the maximum mass increases, the radius of the star at that mass decreases. This is also typical.

\begin{figure}[h]
\includegraphics[width=1.05\columnwidth]{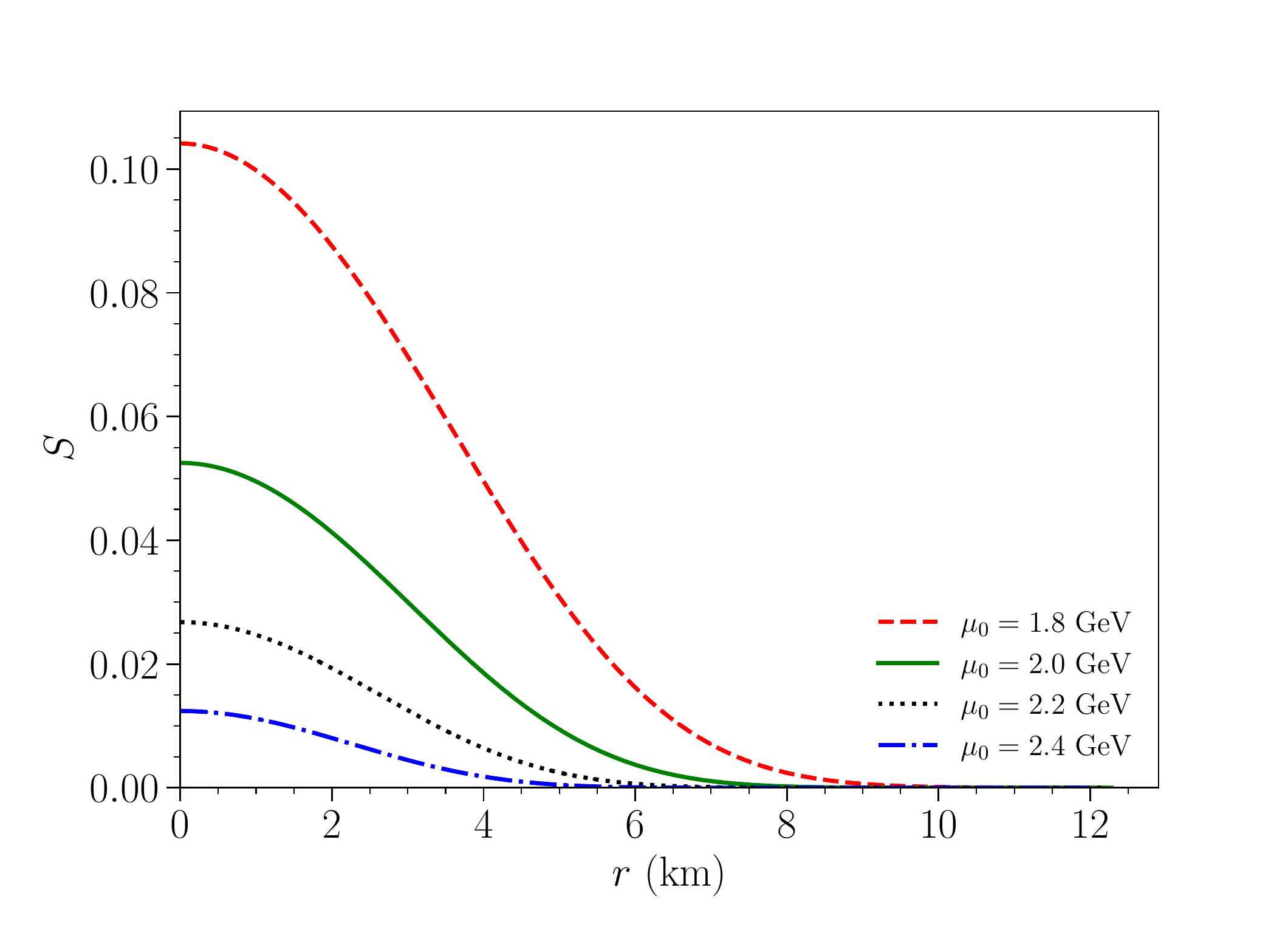}
\caption{Switching function versus radius for the maximum neutron star mass for different choices of $\mu_0$.  It represents the fraction of the total pressure contributed by quark matter.}
\label{Svsr}
\end{figure}

How much quark matter is there in a neutron star? While that seems a natural question, in this case it is ill-posed because the switching function $S$ is a function of $\mu$.  If $n_h$ denotes the baryon density in the hadron phase and $n_q$ denotes the baryon density in the quark phase, then the total baryon density at some value of $\mu$ is
\be
n = \frac{dP}{d\mu} = S n_q + (1-S) n_h + (P_q - P_h) \frac{dS}{d\mu}
\label{Sn}
\ee
which is not just a combination of the the baryon densities in the two phases, but contains an extra term due to the switching function itself.  A better question to ask is what fraction of the pressure is contributed by quarks.  The answer is just $S(\mu)$ which, through Eq. (\ref{Sn}), can also be expressed as $S(n)$.  Figure \ref{Svsr} shows $S$ as a function of radius $r$ for the maximum mass star with a given choice of $\mu_0$.  Between 1 and 10\% of the pressure could be provided by quark matter at the core of the star.

In conclusion, we have studied the masses, radii, and central energy densities of neutron stars using an equation that has a smooth crossover from hadronic matter to quark matter.  This approach is quite flexible in that more sophisticated hadronic equations of state can and should be employed.  This obviously includes the nuclear matter in the crust or outer radius of the star as well as the possibility of including protons, electrons, muons, and hyperons.  What observational effects would support the crossover equation of state to quark matter is an interesting but open question.

\section*{Acknowledgments}
This work was supported by the U.S. DOE Grant No. DE-FG02-87ER40328.

%%%%%%%%%%%%


\begin{thebibliography}{99}
%%%%%%%%%%%%

\bibitem{LV}
B. P. Abbott {\it et al.}, LIGO Scientific Collaboration \& Virgo Collaboration, Phys. Rev. Lett. {\bf 119}, 161101 (2017).

\bibitem{heavies}
F.-K. Thielemann, M. Eichler, I. V. Panov, and B. Wehmeyer, Ann. Rev. Nucl. Part. Sci. {\bf 67}, 253 (2017).

\bibitem{QMseries}
See the proceedings of the Quark Matter Conference series, the most recently available being: Nucl. Phys. A {\bf 1005}, (2021) ed. F. Liu, E. Wang, X.-N. Wang, N. Xu and B.-W. Zhang.

\bibitem{LP1}
J. M. Lattimer and M. Prakash, Phys. Rev. Lett. {\bf 94}, 111101 (2005).

\bibitem{Demorest}
P. B. Demorest, T. Pennucci, S. M. Ransom, M. S. E. Roberts, and J. W. T. Hessels, Nature {\bf 467}, 1081 (2010).

\bibitem{Fonseca}
E. Fonseca {\it et al.}, Astrophys. J. {\bf 832}, 167 (2016).

\bibitem{Antoniadis}
J. Antoniadis {\it et al.}, Science {\bf 340}, 6131 (2013).

\bibitem{Cromartie}
H. T. Cromartie {\it et al.}, Nature Astronomy {\bf 4}, 72 (2020).

\bibitem{Serot-Walecka}
B. D. Serot and J. D. Walecka, Adv. Nucl. Phys. {\bf 16}, 1 (1986); B. D. Serot, Rep. Prog. Phys. {\bf 55}, 1855 (1992);
B. D. Serot and J. D. Walecka, Int. J. Mod. Phys. E {\bf 6}, 515 (1997).

\bibitem{Kapusta-Gale}
J. I. Kapusta and C. Gale, {\it Finite Temperature Field Theory} (Cambridge University Press, 2006).

\bibitem{JEllis}
J. Ellis, J. I. Kapusta, and K. A. Olive, Nucl. Phys. B {\bf 348}, 345 (1991).

\bibitem{SU3}
S. Weissenborn, D. Chatterjee, and J. Schaffner-Bielich, Phys. Rev. C {\bf 85}, 065802 (2012).

\bibitem{latticeQCD}
 Sz. Borsányi, G. Endr\"odi, Z. Fodor, A. Jakovác, S. D. Katz, S. Krieg, C. Ratti,
and K. K. Szabó, J. High Energy Phys. 11 (2010) 077.

\bibitem{matchingpaper}
M. Albright, J. Kapusta, and C. Young, Phys. Rev. C {\bf 90}, 024915 (2014); {\bf 92}, 044904 (2015).

\bibitem{quarkmatter}
 B. A. Freedman and L. D. McLerran, Phys. Rev. D {\bf 16}, 1147 (1977); {\bf 16}, 1169 (1977).

\bibitem{Tews}
I. Tews, J. Carlson, S. Gandolfi, and S. Reddy, Astrophys. J. {\bf 860}, 149 (2018).

\bibitem{quarkyonic1}
L. McLerran and S. Reddy, Phys. Rev. Lett. {\bf 122}, 122701 (2019).

\end{thebibliography}
\end{document}